\documentclass[twocolumn]{aastex701}

\usepackage{booktabs}
\usepackage{amssymb} 
\usepackage{verbatim}

\usepackage{amsmath}

\usepackage{subcaption}
\usepackage{natbib}
\usepackage{placeins}
\begin{document}

\title{Testing the validity of backward orbit reconstruction for outer Local Group dwarf galaxies in IllustrisTNG-100}

\author{Priyanka Verma}
\affiliation{Department of Physics and Astronomy, Mitchell Institute for Fundamental Physics and Astronomy,
Texas A$\&$M University, College Station, Texas 77843, USA}
\email[show]{priyanka\_2943@tamu.edu}

\author{Ekta Patel}
\affiliation{Department of Physics and Astronomy, Villanova University, Philadelphia, PA, USA }
\email{ekta.patel@villanova.edu}

\author{Louis E. Strigari}
\affiliation{Department of Physics and Astronomy, Mitchell Institute for Fundamental Physics and Astronomy,
Texas A$\&$M University, College Station, Texas 77843, USA}
\email{strigari@tamu.edu}

\begin{abstract}
Orbits of Milky Way (MW) and M31 satellites in cosmological simulations have been shown to match analytic orbits computed in fine-tuned models of the extended potential of the MW/M31 halos. Using a sample of Local Group (LG) analogs from the Illustris TNG100 simulations, we investigate whether the same is true for LG satellites near the periphery of the LG turnaround radius ($\sim 1\,{\rm Mpc}$), but outside the virial radii of the MW and M31. From both dark matter-only (TNG100-Dark) and full hydrodynamical (TNG100-1) simulations, we track the orbits of dwarf galaxy analogs and compare them with orbits resulting from an analytic model of the LG potential, which is dominated by the dark matter halos of the MW and M31, along with, where appropriate, their most massive satellites. We find that the reconstructed orbits of the selected outer LG satellites agree with the simulated orbits to within $\lesssim 10\%$ over the last 6 {\rm Gyr}, with the scatter increasing with lookback time and reaching $\sim 40\%$ by 6 {\rm Gyr}. The fact that orbits of outer LG dwarfs track the analytical models suggests that any excess mass in the outer regions of our LG analogs or the tidal field surrounding the LG does not affect the orbits of outer LG satellites. Residual differences between the analytic and simulated orbits are likely due to a combination of mass accretion for the satellites and MW/M31, the non-spherical nature of the MW/M31 potential, and the prescription for dynamical friction in the analytical models.
\end{abstract}


\section{Introduction}
\label{sec:intro}

The Local Group (LG) of galaxies provides a unique laboratory for understanding galaxy formation and evolution in a cosmological context, and for testing the nature of dark matter. Defined as the set of galaxies gravitationally bound to the Milky Way (MW)–Andromeda (M31) barycenter~\citep{1998ARA&A..36..435M,2012AJ....144....4M}, the LG occupies an approximately spherical volume with radius of order $\sim1\,\mathrm{Mpc}$. The total mass of the LG is estimated to be $(2$--$4)\times10^{12}\,{\rm M}_\odot$~\citep{2008MNRAS.384.1459L,2013ApJ...775..102P,2014MNRAS.443.1688D,2014MNRAS.443.2204P,2022MNRAS.511.6193H,2023ApJ...942...18C}, and appears to be largely concentrated in the dark matter halos of the MW and M31~\citep{2022ApJ...928L...5B,2023MNRAS.521.4863S}. Although dozens of lower-mass dwarf galaxies populate the LG and dominate it by number~\citep{2025OJAp....8E.142P}, they contribute only a minor fraction of the total LG mass.

New kinematic measurements, in particular proper motions from Gaia~\citep{2018A&A...616A...1G} and the Hubble Space Telescope (HST)~\citep{2012ApJ...753....8V,2012ApJ...753....7S,2017ApJ...849...93S,2020ApJ...901...43S,2025ApJ...993..228B}, now provide unprecedentedly precise constraints on the motions of LG galaxies. With 6D phase-space information, it is possible to infer orbital histories for many LG members. For example, the MW–M31 relative motion shows evidence for a non-negligible tangential component~\citep{2019ApJ...872...24V,2021MNRAS.507.2592S,2025A&A...701A.265W}, requiring a reassessment of the classic timing argument~\citep{1959ApJ...130..705K}, which assumes a purely radial orbit~\citep{2023ApJ...942...18C}. These results also challenge the simplifying assumption of a head-on MW–M31 collision~\citep{2025NatAs...9.1206S}, motivating more realistic orbital models~\citep[e.g.,][]{2025MNRAS.539..160H}. Accurately modeling the MW--M31 orbit furthermore requires including additional physical effects beyond the simplest two-body approximation, such as the influence of massive satellites~\citep{2019ApJ...872...24V}, the cosmological constant~\citep{2013MNRAS.436L..45P,2014MNRAS.443.2204P}, and may provide a unique probe of local cosmological parameters~\citep{2023ApJ...953L...2B,2025ARA&A..63..431S} 

For MW satellite galaxies, 6D constraints have enabled detailed studies of their orbital histories and infall times~\citep[e.g.,][]{2018ApJ...863...89S,2018A&A...619A.103F,2022A&A...657A..54B}. These data have also been used to place dynamical constraints on the MW mass profile via the phase-space structure of the satellite population~\citep{2018ApJ...857...78P,Callingham2019,2025arXiv250925362H}, and to identify coherent infall and dynamical substructure that encode the MW recent accretion history~\citep{2008MNRAS.385.1365L,2015MNRAS.448L..77D,2020ApJ...893..121P}. In addition, the recent infall of the LMC can measurably perturb the MW’s gravitational potential and barycentric frame, imprinting signatures on satellite orbits and on inferences of the MW mass from satellite kinematics~\citep[e.g.,][]{2013ApJ...764..161K,2015ApJ...802..128G,Erkal2019}. Similar progress is emerging for the M31 satellite system, where HST proper motions are beginning to constrain the orbits of M31 companions and probe possible past interactions~\citep[e.g.,][]{2025ApJ...985..121P}.

In addition to the satellites of the MW and M31, the LG contains a population of {\it outer LG galaxies}: dwarf galaxies within the LG volume but outside the virial radii of both the MW and M31, and not classified as satellites of either host. From a dynamical standpoint, these galaxies inhabit the transition region between the host halos and the LG boundary, and therefore probe the LG mass distribution on intermediate scales; their trajectories are particularly sensitive to the global LG potential and to environmental tides from the surrounding Local Volume. This sensitivity is expected because the LG is embedded in the cosmic web, and LG analogs preferentially inhabit sheet/filament environments in which the surrounding tidal (or velocity-shear) field can influence galaxy motions on Mpc scales \citep{2015ApJ...799...45F,Wang:2026ajj}. In such settings, anisotropies—including lopsided dwarf-galaxy distributions—can arise from the geometry of the MW--M31 pair and from anisotropic accretion, linking the observed phase-space structure of dwarfs to the matter distribution in and around the LG \citep{2016ApJ...830..121L,2022ApJ...938..101S,2024MNRAS.532.2490S}. 

Proper motions from Gaia and HST are currently available for several outer-LG dwarfs \citep{2022A&A...657A..54B}. These measurements enable orbit estimates and can help distinguish first-infall systems from ``backsplash'' galaxies that have previously passed through the virial region of one of the main hosts \citep{2024ApJ...971...98B,2025ApJ...993..228B}.

Interpreting these kinematic measurements requires theoretical models that account for realistic assembly histories and environments. Two complementary simulation approaches are commonly used: (i) unconstrained zoom-in simulations of LG-like systems (e.g., APOSTLE and ELVIS), which provide statistical predictions for satellite populations and their orbits~\citep{2014MNRAS.438.2578G,2016MNRAS.457..844F}; and (ii) constrained simulations of the Local Volume, which aim to reproduce the observed large-scale structure and velocity field around the LG and therefore provide a more direct connection between LG dynamics and the surrounding matter distribution~\citep{2016MNRAS.460L...5C,Sawala:2021npe,McAlpine:2022art}. Recent work using such approaches infers the mass distribution in and around the LG~\citep{Wempe:2024rfj,2026NatAs..10..548W}, including matter beyond the virial radii of the MW and M31 that can contribute to the forces experienced by outer-LG dwarfs.

A common approach to interpret 6D kinematics is to integrate orbits backward in time in semi-analytic potentials~\citep[e.g.,][]{2017MNRAS.464.3825P,2020ApJ...901...43S,2022ApJ...940..136P,2024ApJ...971...98B,2025ApJ...993..228B}. These methods typically assume static, spherical halo potentials and may require modification in the presence of significant time-dependent mass growth, mergers, halo shape evolution, and/or environmental tides~\citep{2022MNRAS.512..739D}. These assumptions can be particularly important for outer-LG dwarfs, whose dynamics can be influenced by the evolving MW/M31 masses and barycentric motion, the time-dependent quadrupole field of the MW--M31 pair, and external tides from large-scale structure. While some studies find good agreement between analytic models and full simulations for the MW satellite population~\citep{2024MNRAS.527.8841S}, the resulting systematic uncertainties have not yet been quantified across the LG as a whole, especially in the regime of outer-LG dwarfs where classifying galaxies as first-infall versus backsplash can hinge on small differences in inferred pericenter times and distances.

In this paper, we systematically investigate the accuracy of analytic orbit reconstruction methods for outer LG galaxies. We use the Illustris TNG100 cosmological simulation~\citep{2019ComAC...6....2N,2018MNRAS.475..648P,2018MNRAS.475..676S,2018MNRAS.475..624N,2018MNRAS.477.1206N,2018MNRAS.480.5113M} to examine the evolution of LG analogs. We compare true orbital trajectories extracted directly from the simulation to trajectories reconstructed by integrating present-day phase-space coordinates backward in time using \texttt{gala}~\citep{2017JOSS....2..388P}, similar to \citet{2024ApJ...971...98B,2025ApJ...993..228B}. 
 
This paper is structured as follows. In Section~\ref{sec:simulation}, we describe the simulations used in our analysis. In Section~\ref{sec:orbit}, we detail our methodology for orbit reconstruction. We present the reconstruction errors for outer satellites in Section~\ref{sec:results}. We discuss the interpretation of these results and compare with previous work in Section~\ref{sec:discussion}.

\section{Simulations}
\label{sec:simulation}
We utilize publicly available data from the \textsc{IllustrisTNG} project~\citep{2019ComAC...6....2N,2018MNRAS.475..648P,2018MNRAS.475..676S,2018MNRAS.475..624N,2018MNRAS.477.1206N,2018MNRAS.480.5113M}, specifically the TNG100-1 hydrodynamical run and its dark matter-only counterpart, TNG100-1-Dark, which we refer to hereafter as Hydro and Dark, respectively. These simulations follow the co-evolution of dark matter, gas, stars, and supermassive black holes in a periodic box of side length $\sim 100\,\mathrm{Mpc}$, with sufficient resolution to resolve dwarf galaxies and satellites around MW and M31–mass hosts. TNG100-1 implements a galaxy formation model including cooling, star formation, stellar and AGN feedback, while TNG100-1-Dark evolves only the collisionless dark matter component. Using both runs allows us to compare orbital properties in the presence and absence of baryonic physics. For the remainder of this section, we describe how we identify MW–M31 analogs and their satellite systems in the simulations, and how we extract their orbits.

\subsection{Selection of Local Group analogs}
\label{subsec:simcuts}
Our first goal is to identify subhalo pairs at $z=0$ that resemble the MW–M31 system in mass, separation, relative kinematics, and isolation. We begin with the full $z=0$ subhalo catalog and apply a mass cut to select all subhalos whose masses are broadly consistent with MW/M31-scale hosts. Unless further noted, when we refer to halo mass, we are referring to the mass of bound particles identified by the \texttt{SUBFIND} algorithm, i.e., the \texttt{SUBFIND} mass. We require the halo mass to be $0.8 \times 10^{12}\, M_\odot \leq M_{total} \leq 3 \times 10^{12}\, M_\odot$ for each member of a candidate pair, consistent with the mass measurements for the MW~\citep{2008MNRAS.384.1459L,BlandHawthorn2016, Eilers2019, PostiHelmi2019, Watkins2019, Wang:2026ajj} and M31~\citep{2017MNRAS.464.3825P, Kafle2018,Tamm2012,2023ApJ...948..104P}.  

From this mass-selected sample, we then search for pairs whose present-day separation and relative velocities are similar to those of the observed LG. Specifically, we require that the separation at $z=0$, $d$, lies in the range $500\,\mathrm{kpc} < d < 1\,\mathrm{Mpc}$, that the relative radial velocity satisfies $-300\,\mathrm{km\,s^{-1}} < v_{\rm rad} < 0\,\mathrm{km\,s^{-1}}$ so that the pair is on an overall approaching orbit. We further demand that the relative tangential velocity is within the measured range, $0 < v_{\rm tan} < 200\,\mathrm{km\,s^{-1}}$~\citep{2019ApJ...872...24V, Salomon2021}. 
The more massive member in each identified pair is referred to as the \emph{primary} and the less massive as the \emph{secondary}. To ensure that both members are of comparable mass and consistent with the approximate MW-M31 mass ratio, we impose a primary–secondary mass ratio $1.5 < M_{\rm primary} / M_{\rm secondary} < 3$. Finally, we impose an isolation criterion by requiring that there is no third subhalo with total mass $M > 8 \times 10^{11}\, M_\odot$ within $1\,\mathrm{Mpc}$ of the barycenter of the primary–secondary pair. Pairs that satisfy all of these conditions are classified as LG analogs. Our full selection criteria are shown in Table ~\ref{tab:lg_pair_selection}, and the results of our selection are shown in Table~\ref{tab:primary_pairs_combined}. After all cuts, we obtain 23 LG analogs from the combined Dark and Hydro simulations.

\begin{table*}[htbp]
\centering
\caption{Selection criteria for identifying LG analog pairs in the TNG simulations at $z=0$.}
\label{tab:lg_pair_selection}
\small 
\setlength{\tabcolsep}{8pt}
\renewcommand{\arraystretch}{1.3}
\begin{tabular}{ll}
\toprule
\textbf{Selection Criterion} & \textbf{Applied Constraints} \\
\midrule
Host mass & $0.8 \times 10^{12}\,M_\odot \leq M_{total} \leq 3 \times 10^{12}\,M_\odot$ \\
Separation at $z=0$ & $500\,\mathrm{kpc} < d < 1\,\mathrm{Mpc}$ \\
Relative radial velocity & $-300\,\mathrm{km\,s^{-1}} < v_{\rm rad} < 0\,\mathrm{km\,s^{-1}}$ \\
Relative tangential velocity & $0 < v_{\rm tan} < 200\,\mathrm{km\,s^{-1}}$ \\
Primary to secondary mass ratio & $1.5 < M_{\rm primary}/M_{\rm secondary} < 3$ \\
Isolation & No third subhalo with total mass $M > 8 \times 10^{11}\,M_\odot$ \\
          & within $\sim 1\,\mathrm{Mpc}$ of the pair barycenter \\
\bottomrule
\end{tabular}
\end{table*}

We note that our sample includes two types of LG analogs. In the more common case, both the primary and the secondary belong to the same friends-of-friends (FoF) group \citep{2019ComAC...6....2N}. In the second case, the primary and secondary belong to different FoF groups but still satisfy all of the LG selection criteria above. We retain both types of pairs in our sample to maximize statistical significance. In total, we have 4/11 LG analogs in the Hydro simulation and 7/12 in the Dark simulation that are in different FoF groups.

\begin{figure*}[t]
    \centering
    \includegraphics[width=1.0\linewidth]{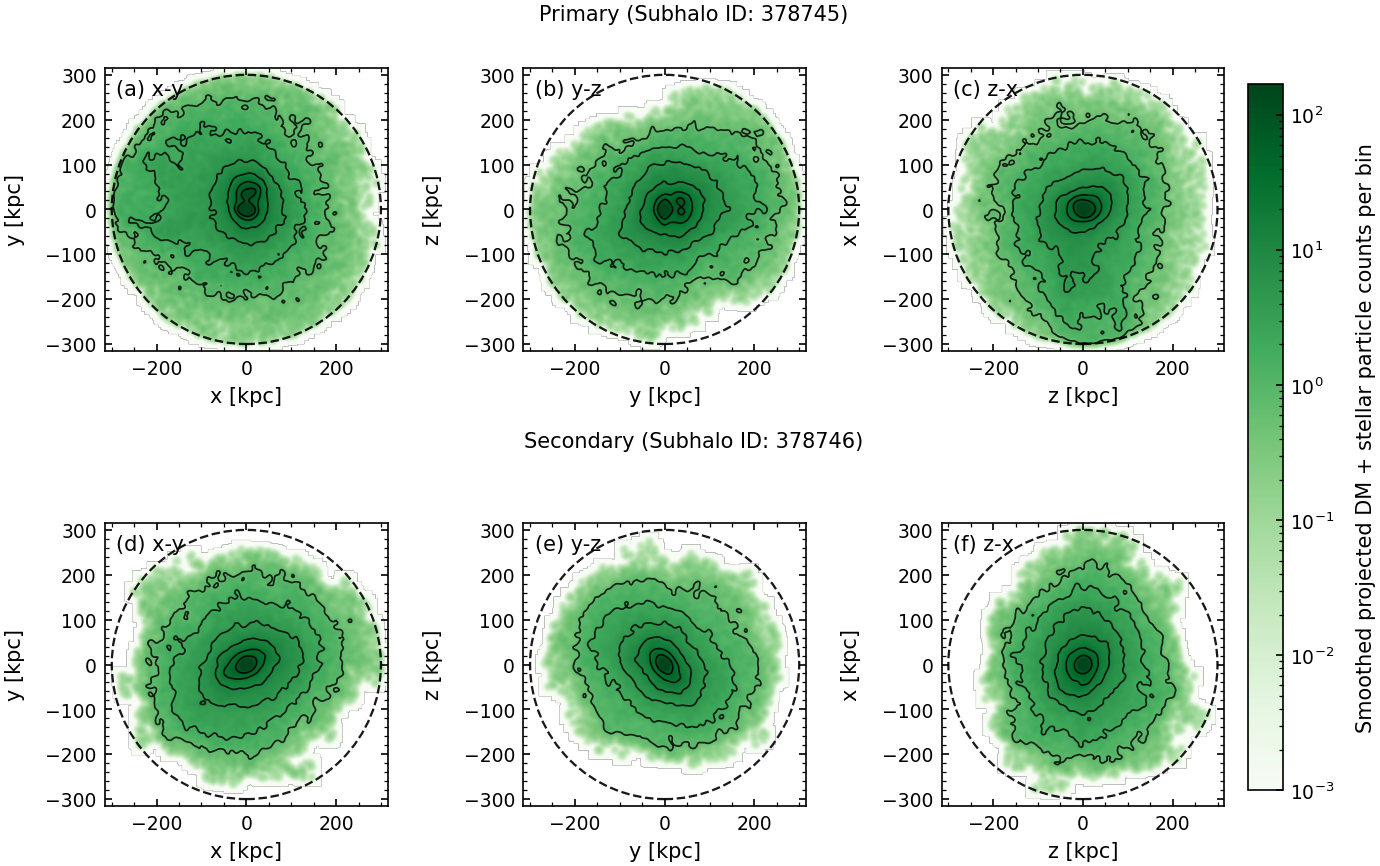}
    \caption{Density projections along the three Cartesian axes for the Hydro LG Pair 378745--378746. The top row is for the primary, and the bottom row is for the secondary. Dark-matter and stellar particles within 300~kpc of each subhalo center are projected onto a two-dimensional grid using random samples of 100,000 dark-matter particles and 100,000 stellar particles, and then Gaussian-smoothed. The color scale and black contours show the smoothed particle counts per bin, while the dashed circles mark a radius of 300~kpc. These figures highlight the non-spherical structure of the host halos in the Hydro sample.}
    \label{fig:HYDRO_Density_1}
\end{figure*}

\begin{figure*}[t]
    \centering
    \includegraphics[width=1.0\linewidth]{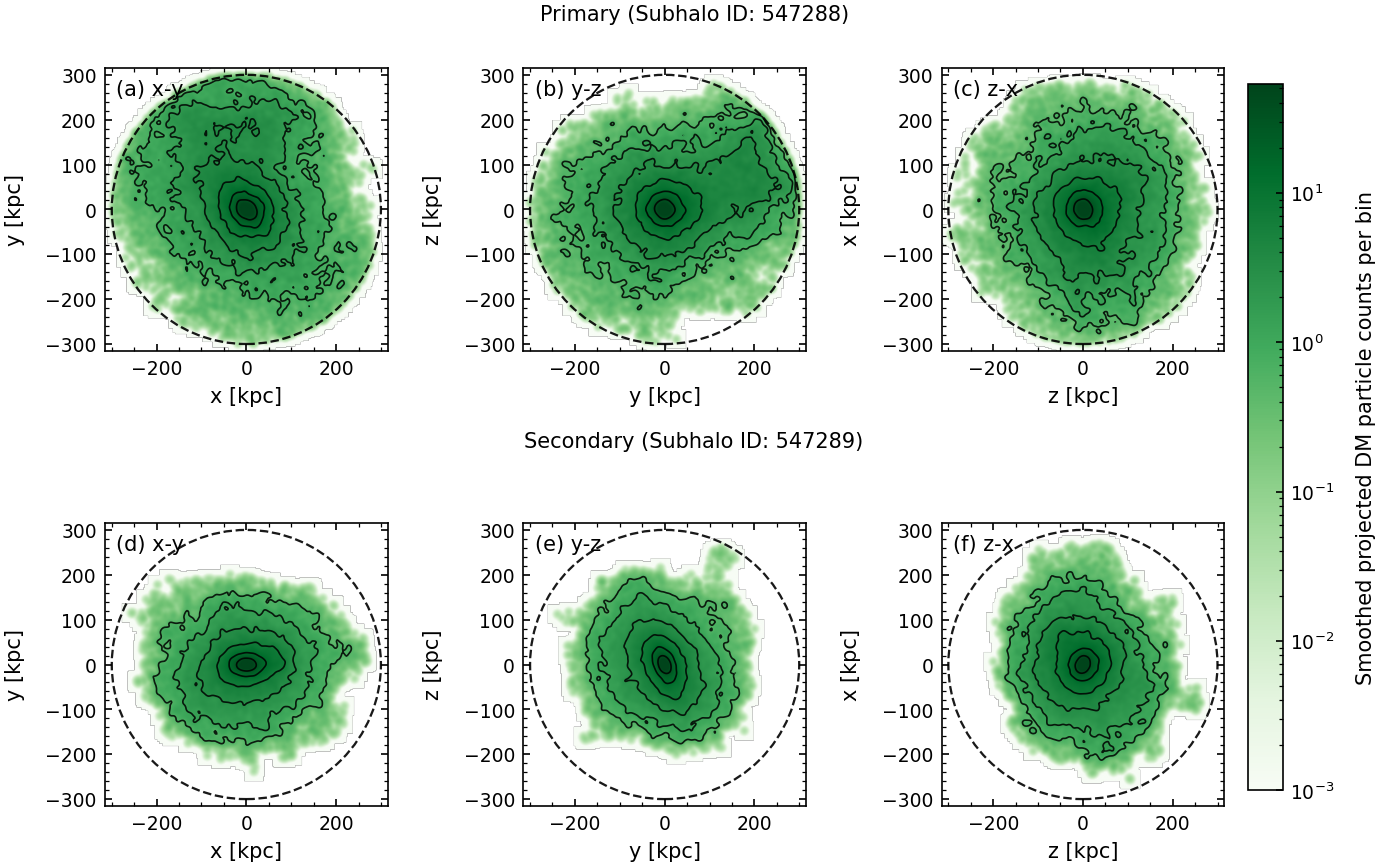}
    \caption{Density projections along the three Cartesian axes for the Dark LG Pair 547288--547289. The top row is for the primary, and the bottom row is for the secondary. Dark-matter particles within 300~kpc of each subhalo center are projected onto a two-dimensional grid using a random sample of 100,000 particles and then Gaussian-smoothed. The color scale and black contours show the smoothed particle counts per bin, while the dashed circles mark a radius of 300~kpc. As in Figure~\ref{fig:HYDRO_Density_1}, this figure illustrates that non-spherical host structure is also present in the Dark sample.} 
    \label{fig:DM_Density_1}
\end{figure*}

Figure~\ref{fig:HYDRO_Density_1} shows two-dimensional density slices for an LG analog in our Hydro sample, and Figure~\ref{fig:DM_Density_1} shows the corresponding slices for an analog from the Dark sample. These figures highlight an overall trend in our sample that halo shapes for primaries and secondaries are uncorrelated. We further note from our sample that there is a range of halo shapes, including cases where the primary is more elongated and others where the halos are more spherical. In some other cases, the inner regions of halos are more elongated and become more spherically distributed near the outer halo.

{\setlength{\tabcolsep}{12pt}
\begin{deluxetable*}{ccccccccc}
\tabletypesize{\scriptsize}
\tablewidth{0pt}
\tablecaption{LG analog pairs in the TNG-100-1 catalogs (Hydro and Dark).
$M_1$ and $M_2$ are halo masses in units of $10^{10}\,M_\odot$; $d$ is the 3D separation in kpc;
$\Delta V_{\rm rad}$ and $\Delta V_{\rm tran}$ are the relative radial and transverse velocities, respectively, in km\,s$^{-1}$.
The final column gives the number of outer satellites, $N$, with total bound \texttt{SUBFIND} mass
$M_{\rm sat}>10^8\,M_\odot$. Outer satellites are farther than 300~kpc from both hosts at $z=0$.
\label{tab:primary_pairs_combined}}
\tablehead{
\colhead{Primary ID} &
\colhead{$M_1$} &
\colhead{Secondary ID} &
\colhead{$M_2$} &
\colhead{$M_1/M_2$} &
\colhead{$d$} &
\colhead{$\Delta V_{\rm rad}$} &
\colhead{$\Delta V_{\rm tran}$} &
\colhead{$N$}
}
\startdata
\multicolumn{9}{c}{\textbf{TNG-100-1 Hydro}} \\
\tableline
378745 & 258.7 & 378746 & 137.9 & 1.9 & 694.0 & -125.0 & 74.2 & 55 \\
387067 & 260.7 & 387068 & 103.2 & 2.5 & 837.7 & -0.6 & 44.1 & 88 \\
407606 & 216.1 & 407607 & 96.3 & 2.2 & 795.4 & -120.8 & 40.5 & 75 \\
413463 & 284.6 & 413464 & 137.4 & 2.1 & 651.3 & -159.8 & 85.0 & 35 \\
417122 & 199.0 & 417123 & 113.5 & 1.8 & 519.4 & -83.4 & 192.8 & 61 \\
420172 & 233.6 & 420173 & 153.5 & 1.5 & 847.5 & -155.0 & 27.9 & 40 \\
434787 & 207.0 & 524216 & 93.9 & 2.2 & 933.1 & -208.0 & 107.9 & 92 \\
449361 & 158.3 & 449362 & 83.9 & 1.9 & 816.9 & -102.6 & 44.2 & 33 \\
449651 & 231.2 & 523425 & 87.2 & 2.7 & 951.3 & -13.7 & 102.5 & 31 \\
457361 & 193.3 & 487769 & 111.8 & 1.7 & 900.8 & -167.7 & 91.4 & 33 \\
466958 & 182.7 & 536830 & 83.3 & 2.2 & 949.6 & -17.3 & 104.2 & 15 \\
\tableline
\multicolumn{9}{c}{\textbf{TNG-100-1 Dark}} \\
\tableline
547288 & 204.3 & 547289 & 128.8 & 1.6 & 591.9 & -88.8 & 33.2 & 130 \\
588227 & 213.5 & 588228 & 138.5 & 1.5 & 577.0 & -134.8 & 81.1 & 48 \\
598806 & 227.9 & 598807 & 98.0 & 2.3 & 784.3 & -136.4 & 44.5 & 74 \\
599434 & 252.0 & 707348 & 116.6 & 2.2 & 945.5 & -201.8 & 105.6 & 125 \\
615377 & 156.6 & 615378 & 88.9 & 1.8 & 573.7 & -128.2 & 57.9 & 30 \\
620503 & 271.4 & 712467 & 104.2 & 2.6 & 904.9 & -5.1 & 124.9 & 60 \\
624556 & 148.9 & 624557 & 80.5 & 1.9 & 535.7 & -43.9 & 141.1 & 40 \\
634354 & 242.4 & 694297 & 115.1 & 2.1 & 868.8 & -158.5 & 98.2 & 58 \\
640569 & 239.3 & 682772 & 149.3 & 1.6 & 943.1 & -130.7 & 68.5 & 53 \\
657694 & 196.1 & 739080 & 81.6 & 2.4 & 745.5 & -105.9 & 73.2 & 32 \\
658251 & 198.7 & 736905 & 94.9 & 2.1 & 947.6 & -22.5 & 102.4 & 20 \\
663267 & 192.4 & 737843 & 80.1 & 2.4 & 926.8 & -82.5 & 100.5 & 27 \\
\enddata
\end{deluxetable*}
}

\subsection{Satellite selection and orbit extraction}
\label{subsec:extraction}

\begin{figure}[h]
    \centering
    \includegraphics[width=0.99\linewidth]{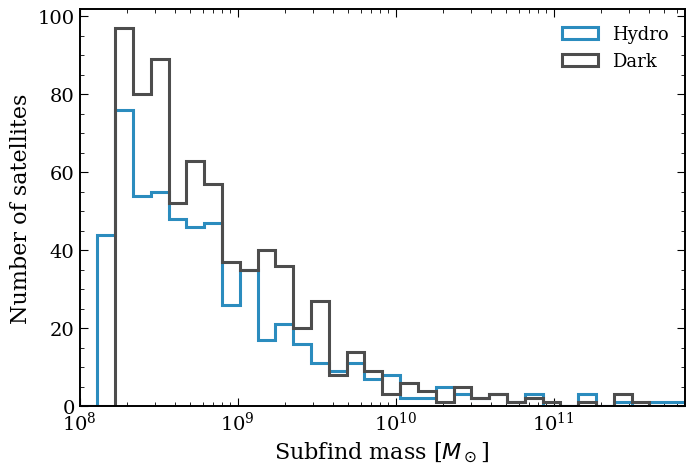}
    \caption{Total \texttt{SUBFIND} mass distributions for the outer LG satellites in Hydro and Dark samples. For Hydro satellites, this mass includes all bound particle/cell types, while in the Dark sample it only includes dark matter particles. The two samples occupy a similar overall mass range, but their abundances differ at fixed mass, particularly at the low-mass end where the Dark sample contains more satellites.}
    \label{fig:mass_hist_Hydro_dark}
\end{figure}

With an LG analog identified, we next select subhalos associated with the FoF group or groups containing the primary and the secondary. If both the primary and the secondary belong to the same FoF group, we search for subhalos within that FoF group. If they occupy two distinct FoF groups, we apply the same satellite selection procedure to each FoF group separately. We utilize the group catalog to identify subhalos within each FoF group, with the primary and secondary hosts excluded from the sample. We select subhalos with total bound \texttt{SUBFIND} mass $M_{\mathrm{total}} > 1 \times 10^{8}\, M_\odot$. This threshold is chosen to eliminate the lowest-mass subhalos near the \texttt{SUBFIND} detection limit, which requires at least 20 gravitationally bound particles/cells \citep{2015MNRAS.449...49R}. The resulting satellite mass distributions for the Hydro and Dark samples are shown in Figure \ref{fig:mass_hist_Hydro_dark}.

\begin{figure*}[t]
    \centering
    \begin{subfigure}[t]{0.49\linewidth}
        \centering
        \includegraphics[width=\linewidth]{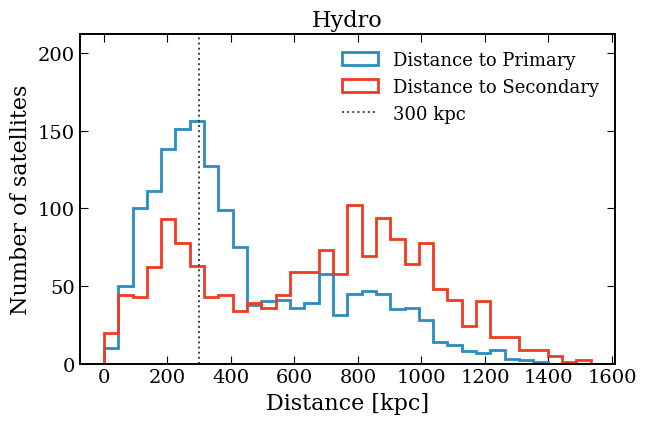}
    \end{subfigure}
    \hfill
    \begin{subfigure}[t]{0.49\linewidth}
        \centering
        \includegraphics[width=\linewidth]{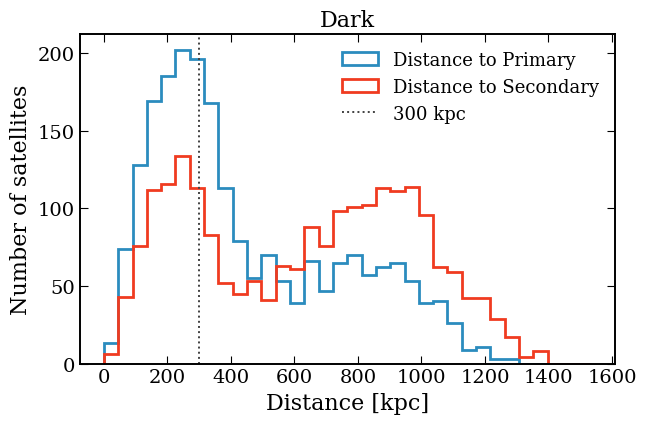}
    \end{subfigure}
    \caption{Distribution of present-day distances between Hydro (left)/Dark (right) satellites and the two hosts in each Local Group analog pair. The blue histogram shows the satellite distances from the primary host, while the red histogram shows the distances from the secondary host. The dashed line at 300 kpc represents our distance cut for satellites. The selected satellites lie to the right of this line for both hosts, ensuring that the sample probes the outer Local Group environment rather than the inner halo of either primary.}
    \label{fig:Distance_Histogram_Plots}
\end{figure*}

We further apply a distance cut to these satellites so that their three-dimensional separation from both the primary and the secondary exceeds 300 {\rm kpc} at z=0. This defines our population of outer LG satellites. The present-day distance distributions relative to the primary and secondary hosts are shown for the Hydro and Dark samples in Figure~\ref{fig:Distance_Histogram_Plots}. 

For each LG analog, we extract satellite orbits directly from the simulation. Starting from the $z=0$ snapshot, we follow the Sublink merger tree of each primary, secondary, and satellite back in time through successive snapshots. At each snapshot, we compute the relative satellite positions and obtain the distances to the primary and secondary as a function of lookback time. We follow the orbits back to a maximum lookback time of $\sim 6\,\mathrm{Gyr}$. This is motivated by similar studies, e.g.~\citet{2020ApJ...893..121P}, which also adopted a 6 Gyr timeline, which is long enough to sample multiple satellite orbits, while limiting the impact of host mass growth and satellite tidal mass loss. Additionally,  previous studies have also shown that beyond this time, the MW and M31 analogs are in a more rapid accretion phase~\citep[]{2023MNRAS.526L..77S}.   

\subsection{Mass accretion histories}
\label{subsec:accretion}

\begin{figure*}[t]
    \centering
    \begin{subfigure}[t]{0.49\linewidth}
        \centering
        \includegraphics[width=\linewidth]{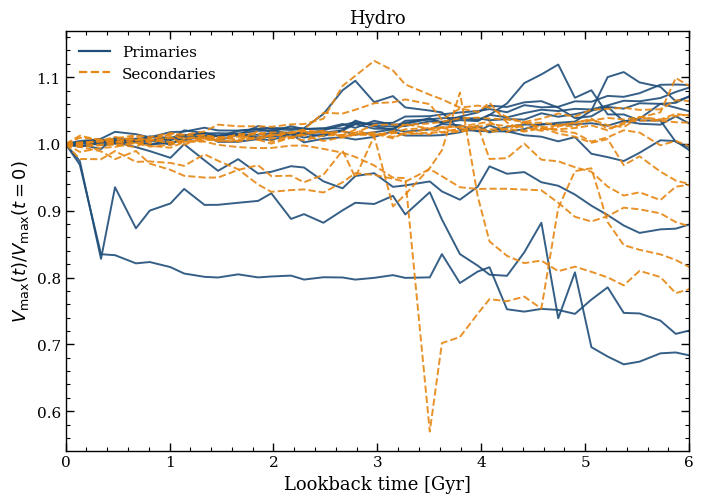}
    \end{subfigure}
    \hfill
    \begin{subfigure}[t]{0.49\linewidth}
        \centering
        \includegraphics[width=\linewidth]{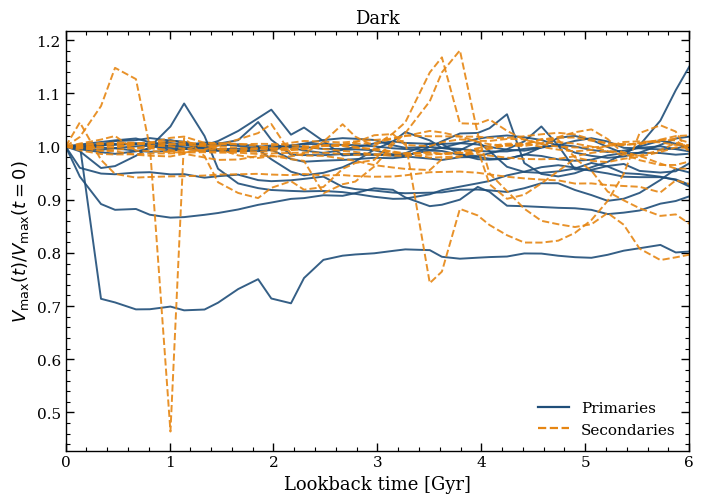}
    \end{subfigure}
    \caption{The maximum circular velocity as a function of lookback time, normalized by the present-day value of the maximum circular velocity, i.e. $V_{\max}(t)/V_{\max}(t=0)$. Plots are for primaries and secondaries in Hydro (left) and Dark (right) LG analogs. This figure shows that the host potentials evolve moderately over the last 6~Gyr, which may contribute to the growing orbit-reconstruction errors at earlier times.}
    \label{fig:Vmax_plots}
\end{figure*}

In addition to the orbital positions and velocities, we track the evolution of the mass accretion history of both the primaries, secondaries, and their satellites as a function of lookback time. We quantify the mass accretion history in terms of the maximum circular velocity, $V_{\max}$. The $V_{\max}(t)$ history serves as a useful proxy for changes in the underlying mass distribution, allowing us to identify episodes of strong tidal stripping, mergers, or rapid mass accretion that may correlate with features in the orbital trajectories. The $V_{\max}$ histories of Hydro and Dark primaries and secondaries are shown in Figure~\ref{fig:Vmax_plots}. A large fraction of our hosts have stable mass accretion histories out to lookback times of 6 Gyr, with at most an $\sim 50\%$ change in the $V_{\max}$ from this time to the present in a few halos. The effect of this magnitude of evolution will be discussed further in Section \ref{subsec:systematics}.

%\FloatBarrier
\section{Orbit Modeling}
\label{sec:orbit}
We perform orbit modeling with the Python package \texttt{gala}~\citep{2017JOSS....2..388P}, using its implementations of gravitational potentials, including \texttt{NFWPotential}, \texttt{MiyamotoNagaiPotential}, and \texttt{PlummerPotential}, together with its orbit integration tools. A custom dynamical friction formula has been implemented in conjunction with \texttt{gala}. In this section, we discuss the details of our orbit modeling, starting by discussing the host potentials, then moving on to the satellite potentials and the satellite orbits.

\subsection{Dark halo model parameters}
We model the primary and secondary in each LG analog as an extended NFW dark matter halo~\citep{1997ApJ...490..493N}. The NFW profile is defined in terms of its scale density and scale radius, the latter of which is defined as $r_s \equiv r_{\rm vir}/c_{\rm vir}$, where $r_{\rm vir}$ is the virial radius and $c_{\rm vir}$ is the concentration. To assign an effective \(r_s\) to each subhalo at \(z=0\), we follow the standard parameter definitions for an NFW halo \citep[e.g., the Appendix of][]{2012ApJ...753....8V}. In order to derive $r_{\rm vir}$, we compute the virial overdensity, $\Delta_{vir}$, using the fitting formula of \citet{1998ApJ...495...80B} for a $\Lambda$CDM cosmology. The virial overdensity is a function of $\Delta_{\rm vir}$, halo mass, and cosmological parameters. For the cosmological parameters we adopt the \textsc{IllustrisTNG} Planck2015 \citep{2016A&A...594A..13P} cosmological parameters \(h=0.6774\) and \(\Omega_m=0.3089\). 

For the halo mass definition, we consider a ``halo-only'' mass estimate by subtracting the stellar component,
$M_{\rm halo} = M_{\rm tot} - M_{\star}$, where $M_{\star}$ is zero for Dark halos. Given this adopted mass, we compute an effective virial radius corresponding to that mass under the spherical-overdensity convention. For the given halo mass, we assign concentrations using the mass--concentration relation of \citet{Dutton_2014}. Then, for each subhalo, we compute
\begin{equation}
\qquad
r_{s,{\rm halo}} = \frac{r_{\rm vir}(M_{\rm halo})}{c_{\rm vir}(M_{\rm halo})}. 
\label{eq:rs_tot_rs_halo}
\end{equation}
Here $r_{s,{\rm halo}}$ uses the halo-only mass estimate, $M_{\rm halo}$.

\subsection{Miyamoto--Nagai disk scale parameters}

For the primary and secondary, we estimate the parameters of the Miyamoto--Nagai disk \citep{1975PASJ...27..533M} using the stellar half-mass radius from the \textsc{IllustrisTNG} catalog which is given in units of \({\rm ckpc}/h\). Assuming an exponential disk profile \citep{1970ApJ...160..811F}, we approximate the radial scale length as $a \simeq \frac{r_{1/2}}{1.678}$, where \(r_{1/2}\) is the stellar half-mass radius at \(z=0\) from the halo catalogs. We further adopt a thin-disk approximation for the vertical scale parameter such that $b \simeq 0.1\,a$. 

This model, consisting of NFW host halos for both Hydro and Dark systems and an additional Miyamoto--Nagai disk component for Hydro hosts, defines our fiducial orbit model. In later sections, we compare this fiducial model to two simplified comparison models: a point-mass host model and, for the Hydro sample, an NFW-only model without a disk component.

\subsection{Satellite potentials and orbits}
For the satellites, we assume spherical Plummer potentials \citep{Plummer1911}, where the Plummer radius, $k_{\mathrm{sat}}$, is obtained from 
\begin{equation}
    M(r_{\mathrm{half}}) =
    \frac{M_{\mathrm{tot}}\, r_{\mathrm{half}}^{3}}
         {\left(r_{\mathrm{half}}^{2} + k_{\mathrm{sat}}^{2}\right)^{3/2}} .
\end{equation}
Here $M_{\mathrm{tot}}$ denotes the total subhalo mass at $z=0$, $M(r_{\mathrm{half}})$ is set to half of this value, and $r_{\mathrm{half}}$ is the half mass radius. We use the $r_{\mathrm{half}}$ values provided for each subhalo in the TNG catalogs. Given $M_{\mathrm{tot}}$ and $r_{\mathrm{half}}$, we solve for $k_{\mathrm{sat}}$.
We use the same Plummer model approximation for satellites in both the Hydro and Dark simulation samples.

\subsection{Orbits integrations} 

With the host and satellite potentials defined, we run the orbit integrations backward in time for \(6\,\mathrm{Gyr}\) to reconstruct the past orbital histories of simulated satellite galaxies. In our calculations, each satellite is integrated independently in the combined gravitational potential of the two hosts, whose positions evolve under their mutual gravitational attraction, while neglecting the gravitational influence of other satellites. This three-body approximation is well justified for the numerous low-mass dwarf satellites in the LG \citep{2022MNRAS.512..739D}. 

Some of our LG analogs contain massive satellite galaxies that may affect the orbits of lower mass satellites. This is akin to the impact of the LMC in the observed LG, which displaces the MW from the pair barycenter and perturbs the orbits of other satellites \citep{2015ApJ...802..128G, Garavito-Camargo2019, Erkal2019, 2020ApJ...893..121P}. A similar but weaker effect is expected for the M33 and M31 case~\citep{2025ApJ...985..121P}. In these cases with massive satellites, we switch to more than three-body (typically four- or five-body) integrations in \texttt{gala}, treating the MW, M31, the massive satellite(s), and the satellite of interest as mutually interacting bodies. For the massive satellites, we assume Plummer potentials as described above and in ~\citet{2025ApJ...993..228B}. 

For the satellites, we implement dynamical friction using the Chandrasekhar formula \citep{1943ApJ....97..255C}.
We adopt a mass-dependent Coulomb logarithm, motivated by \citet{2020ApJ...893..121P}. For lower mass satellites, we use the distance-dependent form of \citet{2003ApJ...582..196H}, \(\ln\Lambda=\ln[r/(1.4a_{sat})]\) where \(r\) is the satellite-host separation and \(a_{\rm sat}\) is the Plummer scale radius used in our satellite modeling. For massive satellites, we use the parametrized form of \citet{2012ApJ...753....8V}, \(\ln\Lambda=\max[L,\alpha\ln(r/Ca_{\rm sat})]\), with \(L=0.02\), \(C=0.17\), and \(\alpha=0.15\). This mass-dependent approach accounts for the stronger dynamical friction expected for massive satellites relative to lower-mass systems.

\section{Results}
\label{sec:results}

We now compare the satellite trajectories extracted directly from the TNG simulations to the trajectories reconstructed with the \texttt{gala} orbit integrations described in Section~\ref{sec:orbit}. Our goal is to quantify how accurately present-day phase-space measurements recover the past orbital evolution of outer satellites in LG analog systems when the gravitational potential is modeled using only the dominant host galaxies.

\subsection{Orbit reconstruction} 

For each satellite, we compute its distance from both the primary and secondary hosts as a function of lookback time. We then compare the reconstructed and simulated distances using the signed fractional distance error
\begin{equation}
    \eta(t_{\rm lb}) =
    \frac{r_{\rm model}(t_{\rm lb}) - r_{\rm sim}(t_{\rm lb})}
         {r_{\rm sim}(t_{\rm lb})},
    \label{eq:eta}
\end{equation}
where \(r_{\rm model}\) is the distance obtained from the \texttt{gala} orbit integration, \(r_{\rm sim}\) is the corresponding distance measured from the simulation merger tree, and \(t_{\rm lb}\) is the lookback time. Positive values of \(\eta\) indicate that the reconstructed orbit places the satellite at a larger distance from the host than in the simulation, while negative values indicate that the reconstructed orbit places the satellite analog closer to the host.

\begin{figure*}[t]
    \centering
    \begin{subfigure}[t]{0.49\linewidth}
        \centering
        \includegraphics[width=\linewidth]{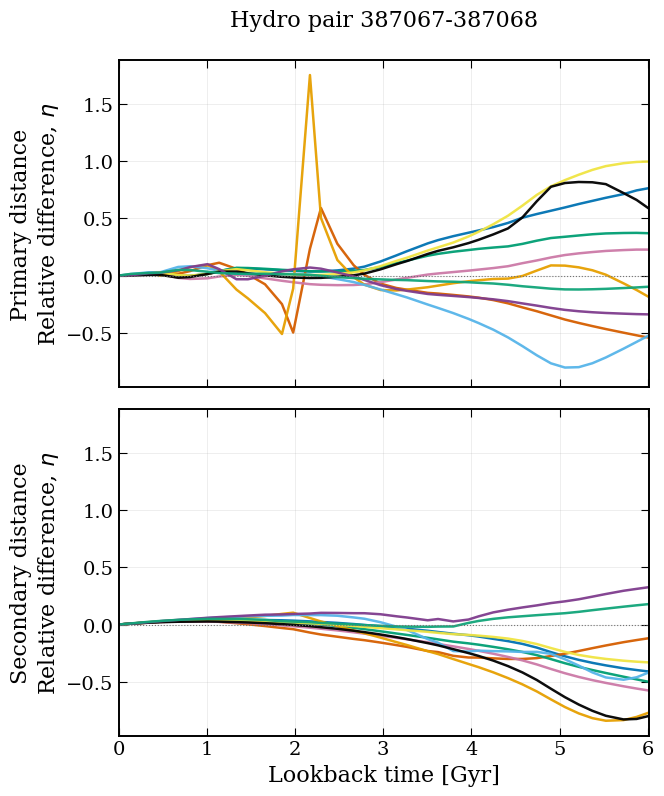}
    \end{subfigure}
    \hfill
    \begin{subfigure}[t]{0.49\linewidth}
        \centering
        \includegraphics[width=\linewidth]{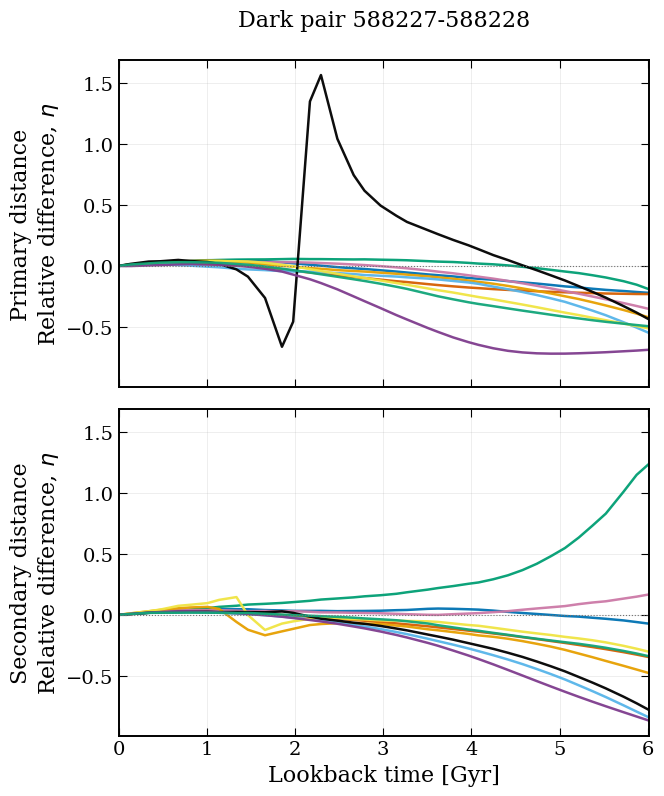}
    \end{subfigure}
    \caption{Relative difference $\eta$ between the \texttt{gala} reconstructed orbits and the TNG simulation for the ten most massive outer satellites in the Hydro pair 387067--387068 and Dark pair 588227--588228. For the Hydro pair, the primary and secondary masses are \(2.607 \times 10^{12}\,M_\odot\) and \(1.032 \times 10^{12}\,M_\odot\), respectively. For the Dark pair, the primary and secondary masses are \(2.135 \times 10^{12}\,M_\odot\) and \(1.385 \times 10^{12}\,M_\odot\), respectively. The top and bottom panels show $\eta$ computed using satellite distances from the primary and secondary, respectively. The reconstructions agree best near (z=0) and diverge toward earlier lookback times.}
    \label{fig:TNG_vs_GALA_orbits}
\end{figure*}

Figure~\ref{fig:TNG_vs_GALA_orbits} shows examples of the orbit reconstructions for the ten most massive outer satellites in one hydro LG analog and one dark LG analog. In both cases, the reconstructed and simulated trajectories agree well near \(z=0\), as expected because the \texttt{gala} integrations are initialized from the present-day phase-space coordinates in the simulations. At larger lookback times, the reconstructed and simulated trajectories gradually diverge. This divergence is not monotonic for every individual satellite, because the error depends on orbital phase, close passages, and the relative configuration of the two hosts. Nevertheless, the general trend is clear: backward orbit reconstruction becomes less accurate as the integration is extended further into the past. This behavior is consistent with previous studies showing that orbit reconstruction in static or idealized potentials becomes increasingly uncertain at earlier lookback times \citep{2024MNRAS.527.8841S,2022MNRAS.512..739D}.

\begin{figure*}[t]
    \centering
    \includegraphics[width=1\linewidth]{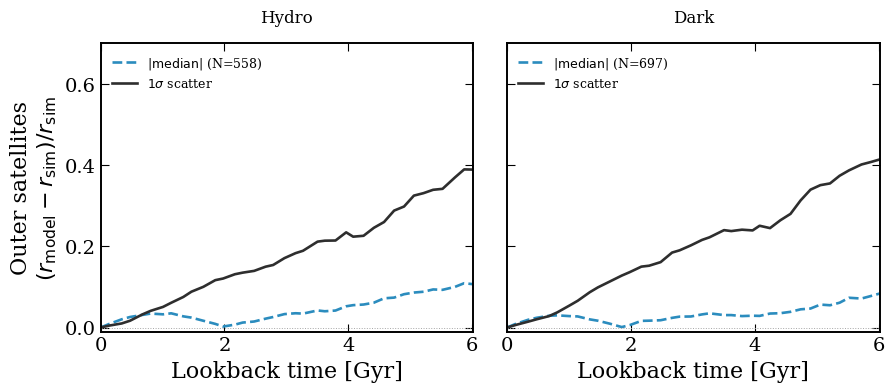}
    \caption{Fractional errors between the reconstructed and simulated orbits as a function of lookback time for the fiducial orbit model. The hosts are modeled with NFW dark matter halos, and Hydro hosts additionally include the disk component described in Section~\ref{sec:orbit}. The curves show the absolute value of the median signed fractional distance error, while the shaded regions show the \(1\sigma\) scatter, defined as half of the 16th--84th percentile range. The small median errors indicate little systematic bias, while the increasing scatter shows growing uncertainty for individual orbits.}
    \label{fig:Santisteven_direct_comparison_plot}
\end{figure*}

To quantify this behavior over the full satellite population, Figure~\ref{fig:Santisteven_direct_comparison_plot} shows how the orbit distributions compare as a function of lookback time for our fiducial orbit model. For each satellite, we compute \(\eta\) with respect to the host that is closer to the satellite at \(z=0\). This choice provides a single representative distance error for each satellite and allows a direct comparison with the convention adopted by \citet{2024MNRAS.527.8841S}. At each lookback time, we summarize the population by the absolute value of the median signed fractional error, \(|{\rm median}(\eta)|\), and by the \(1\sigma\) scatter, defined as half of the 16th--84th percentile range. The median measures the typical bias of the reconstructed distances relative to the simulated distances, while the scatter measures the satellite-to-satellite variation in reconstruction accuracy. Note that a small median does not imply that every individual orbit is accurately recovered; rather, it indicates that the reconstructed distances are not systematically too large or too small for the population as a whole. The scatter provides a more useful estimate of the uncertainty for individual satellites.

For the fiducial orbit model, the median fractional error remains small over the full integration interval, while the scatter grows steadily with lookback time. This indicates that the simplified two-host model captures the mean behavior of the satellite population, but that individual reconstructed orbits become increasingly uncertain at earlier times. By a lookback time of $\sim 6\,\mathrm{Gyr}$, the scatter reaches approximately \(\sim 0.4\), implying that individual satellite distances can differ from the simulated values by up to 40\% even though the population median remains close to zero.

The relatively small median errors in Figure~\ref{fig:Santisteven_direct_comparison_plot} suggest that the dynamics of outer LG satellites are dominated by the two most massive halos in the system. If diffuse matter, additional substructure, or external tides dominated their dynamics over the last several Gyr, then a model including only the two main hosts would be expected to produce a larger systematic offset. Instead, the fiducial orbit model successfully captures the broad radial evolution of the satellite population, although with increasing object-to-object scatter at earlier times.

\subsection{Comparison to point mass potential} 

\begin{figure*}[t]
    \centering
    \includegraphics[width=1\linewidth]{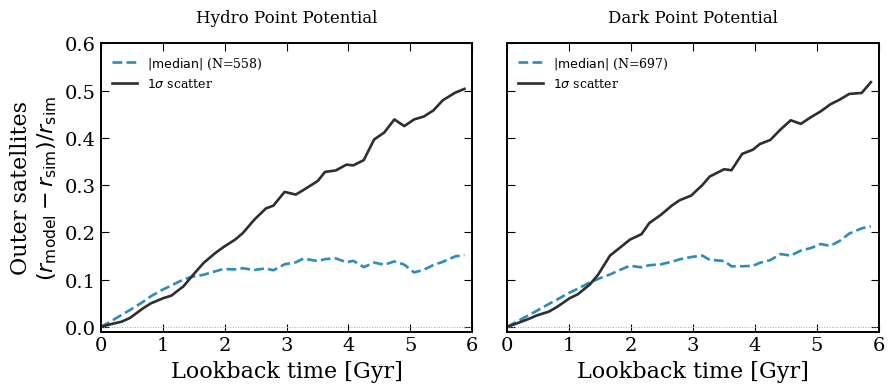}
    \caption{Fractional errors between the reconstructed and simulated orbits as a function of lookback time when the primary and secondary are modeled as point-mass potentials. The point-potential model produces larger median errors and larger scatter than the fiducial orbit model (Figure~\ref{fig:Santisteven_direct_comparison_plot}), indicating that the internal mass distribution of the hosts is important for accurate orbit reconstruction.}
    \label{fig:Santisteven_direct_comparison_plot_point_potential}
\end{figure*}

We next test how sensitive the reconstruction is to the assumed form of the host potential. As an extreme limiting case, we replace the fiducial extended NFW (NFW-plus-disk) model with point-mass potentials for the primary and secondary, assigning each host its present-day \texttt{SUBFIND} mass. The results for fractional errors are shown in Figure~\ref{fig:Santisteven_direct_comparison_plot_point_potential}. The point-mass model follows the same qualitative trend as the extended model: the median error remains modest, while the scatter increases with lookback time. However, the point-mass model produces systematically larger median errors and a larger scatter. By $\sim 6\,\mathrm{Gyr}$ ago, the scatter reaches approximately \(\sim 0.5\), compared with \(\sim 0.4\) for the extended-potential model.

This comparison shows that the detailed radial structure of the host potentials, modeled as extended halo-plus-disk potentials, improves the recovery of individual satellite trajectories. The difference between the point-mass and extended-potential models is not so large as to change the qualitative conclusion that the two main hosts dominate the dynamics, but it is large enough to matter for more detailed orbit reconstruction. In particular, the point-mass approximation is less reliable for satellites that have passed relatively close to one of the hosts, where the enclosed mass profile and force law differ most strongly from those of an extended halo.

\subsection{Modeling baryons} 

\begin{figure}[t]
    \centering
\includegraphics[width=0.99\linewidth]{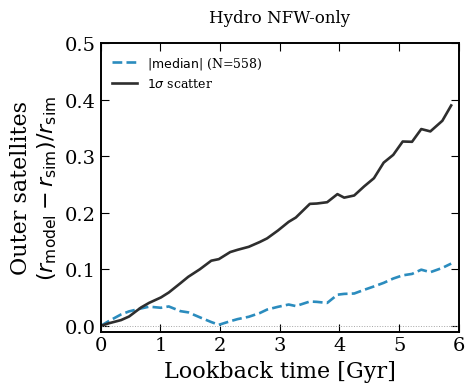}
    \caption{Fractional errors between the reconstructed and simulated orbits as a function of lookback time for the Hydro Local Group analogs when the primary and secondary are modeled with NFW halos only, without a disk component. The similarity to the Hydro fiducial orbit model indicates that the approximate disk component does not dominate the reconstruction error for the outer-satellite population.}
    \label{fig:Santisteven_direct_comparison_plot_hydro_NFW_only}
\end{figure}

Finally, we test the impact of the baryonic disk component in the Hydro sample. Figure~\ref{fig:Santisteven_direct_comparison_plot_hydro_NFW_only} shows the Hydro reconstruction errors when the primary and secondary are modeled using only NFW halos, with no disk component. The resulting median errors and scatter are similar to those obtained when the disk is included. This indicates that, for the outer satellites considered here, the approximate disk component is not the dominant factor setting the reconstruction accuracy.

This result should not be interpreted as evidence that baryons are dynamically unimportant in all contexts. Disk components can strongly affect and even disrupt satellites on small-pericenter orbits, especially within the inner halo \citep{2017MNRAS.471.1709G,2019MNRAS.487.4409K}. However, our sample is selected to contain satellites outside \(300\,\mathrm{kpc}\) from both hosts at \(z=0\), and the statistic in Figure~\ref{fig:Santisteven_direct_comparison_plot_hydro_NFW_only} summarizes the population as a whole. For this outer-satellite population, the reconstruction errors appear to be driven primarily by other effects, such as host growth, deviations from spherical symmetry, and tidal stripping resulting from close passages with the hosts.

Overall, the results in this section show that backward integrations provide a useful statistical description of outer LG satellite orbits, particularly at recent lookback times. The small median errors indicate that the two dominant hosts capture the leading contribution to the satellite dynamics. However, the steadily increasing scatter demonstrates that individual satellite histories become progressively less certain toward earlier times. The extended-potential model performs better than the point-potential model, while the inclusion of an approximate disk component has little effect on the outer-satellite population.

\section{Discussion}
\label{sec:discussion}

The results in Section~\ref{sec:results} show that backward orbit reconstruction in LG potentials recovers the broad evolution of outer satellites, but with errors that grow systematically with lookback time. This behavior is expected: the reconstructed and simulated orbits are constrained to agree at \(z=0\), but the backward integrations neglect several aspects of the true cosmological evolution of the system. In this section, we discuss the main physical effects that may contribute to the residual differences between the reconstructed and simulated trajectories, and place our results in the context of previous orbit reconstruction studies.

\begin{figure*}[t]
\centering
\begin{subfigure}[t]{0.48\linewidth}
    \centering
    \includegraphics[width=\linewidth]{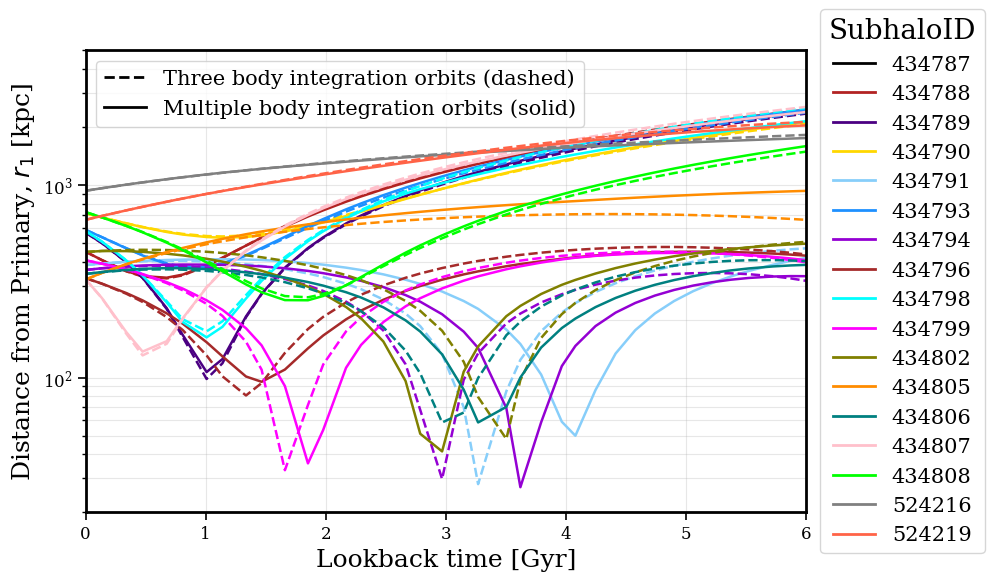}
    \label{fig:multi_vs_three_r1}
\end{subfigure}
\hfill
\begin{subfigure}[t]{0.48\linewidth}
    \centering
    \includegraphics[width=\linewidth]{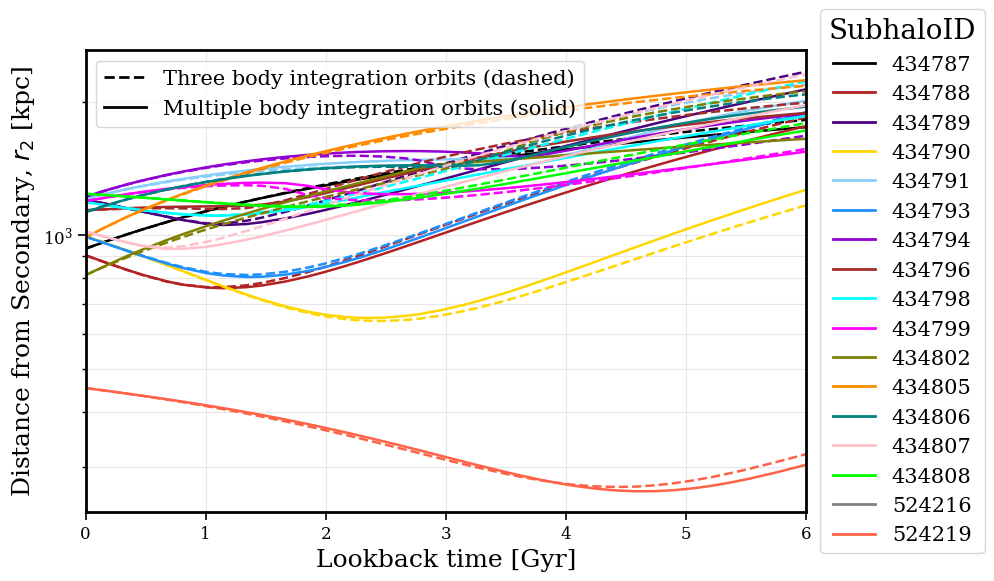}
    \label{fig:multi_vs_three_r2}
\end{subfigure}
\caption{Comparison of satellite orbits computed from multiple-body integrations (solid) and three-body integrations (dashed) as a function of lookback time. This shows that massive satellites can affect individual systems, but they do not dominate the population-level reconstruction error.}
\label{fig:multi_vs_three_r1r2}
\end{figure*}

\subsection{Systematics in backward integrations} 
\label{subsec:systematics}
First, we address the assumption that the host potentials are static. In the \texttt{gala} integrations, the primary and secondary halos are assigned their present-day masses and scale radii, and these properties are held fixed over the full \(6\,\mathrm{Gyr}\) integration interval. In the simulations, however, the hosts continue to grow and evolve. Figure~\ref{fig:Vmax_plots} shows that many of the hosts have relatively stable \(V_{\max}\) histories over the last \(6\,\mathrm{Gyr}\), with changes of order tens of percent, but a few systems exhibit more pronounced or abrupt evolution. Such evolution changes both the depth and shape of the potential experienced by satellites.

Recent merger activity may also play a role. M31 appears to have experienced a significant merger event approximately \(\sim 2-3\,\mathrm{Gyr}\) ago \citep{2018NatAs...2..737D,2018MNRAS.475.2754H}, indicating that similarly recent events are plausible in LG-like environments. In cosmological simulations, however, the merger histories of LG analogs depend sensitively on the imposed isolation criteria. For example, isolated LG analogs have a reduced rate of recent major mergers, with no major mergers since \(z=0.3\) in their selected sample~\citep{2025MNRAS.539..160H}. Our analogs are selected to be broadly isolated, but they are still drawn from a fully cosmological environment and therefore retain a range of accretion histories. This diversity likely contributes to the system-to-system variation in reconstruction accuracy, but an in-depth analysis of how accretion histories can affect orbit reconstruction is beyond the scope of this work.

Another important approximation is the geometry of the host potentials. Our fiducial orbit model represents the primary and secondary dark matter halos as spherical NFW profiles, with an additional approximate disk component for the hydrodynamical hosts. In contrast, the simulated halos are not perfectly spherical. The density maps in Figures~\ref{fig:HYDRO_Density_1}--\ref{fig:DM_Density_1} show that the primaries and secondaries span a range of shapes, including visibly elongated and asymmetric cases. Satellites orbiting in such potentials can experience torques and orbital precession that are absent in spherical models. These effects should be most important for satellites that pass relatively close to one of the hosts, but they can also affect outer satellites over several Gyr through the accumulated influence of the non-spherical mass distribution.

Close passages provide another source of sensitivity. During pericentric passages, satellite trajectories depend strongly on the detailed mass profile of the host and on the instantaneous configuration of the LG analog. Small differences in the assumed potential, host position, or host velocity can therefore lead to substantial differences in the recovered orbit after the passage. Furthermore, satellites may experience tidal stripping during close passages, and this is not accounted for in our backward orbit reconstructions. This sensitivity may help explain why the scatter increases toward larger lookback times even when the median bias remains modest. It also implies that quantities tied to individual orbital phases, such as pericenter time, pericenter distance, or backsplash classification, may be more uncertain than the broad radial evolution of the satellite population.

\subsection{Impact of massive satellites} 

We have also tested whether massive satellites within the LG analogs are a dominant source of the mismatch. Massive companions can perturb the barycentric motion of the host pair and alter the trajectories of other satellites, analogous to the effect of the LMC on the MW satellite system \citep{2015ApJ...802..128G,Garavito-Camargo2019,Erkal2019,2022ApJ...940..136P} and, to a lesser extent, the possible influence of M33 on M31 satellites \citep{2025ApJ...985..121P}. Table~\ref{tab:massiveSatellites} lists the systems in our sample that contain satellites with masses exceeding ten percent of the less massive host. For these cases, we compare the default three-body integrations to integrations that include the additional massive bodies. Figure~\ref{fig:multi_vs_three_r1r2} shows that the resulting trajectories are often similar over the full integration interval, even after accounting for perturbations from massive satellites and barycentric evolution. Figure~\ref{fig:multi_vs_three_r1r2} shows that the satellites most affected by the contribution of massive satellites are those that have close passages relative to the primary. In these cases, the impact of the massive satellites is to shift pericenter timing to later lookback times and, in some cases, to change the pericenter distance.

In the LG, it has been shown that contributions from massive satellites crucially depend on their orbital trajectories (i.e., the MW-LMC orbital configuration is vastly different from that of M31-M33). Thus, while massive satellites can be important in the evolution of individual systems, as is the case with the real LG, they do not appear to be the dominant driver of the overall reconstruction error across the full sample of outer satellites.

\subsection{Comparison with previous work}

Previous studies have tested backward orbit reconstruction for satellites of MW-mass hosts. Our analysis extends this problem to LG analog pairs, where satellites evolve in the combined potential of two massive galaxies and their shared environment. Table~\ref{tab:model_comparison} summarizes the main differences between our study and two closely related works, including the simulation suite, satellite selection, host environment, and orbit model.

A particularly useful comparison is \citet{2024MNRAS.527.8841S}, who reconstructed satellite orbits in FIRE-2 MW/M31-mass analogs. They quantified the absolute value of the median signed fractional distance error and the \(1\sigma\) scatter, defined as the half-width of the 16th--84th percentile range. We adopt the same convention in Figures~\ref{fig:Santisteven_direct_comparison_plot}, \ref{fig:Santisteven_direct_comparison_plot_point_potential}, and \ref{fig:Santisteven_direct_comparison_plot_hydro_NFW_only}. Although our host models are simpler and our systems contain two massive hosts rather than a single isolated host, the scale and time dependence of the reconstruction errors are broadly similar. In both analyses, reconstructed orbits agree best at recent times and become increasingly uncertain toward earlier lookback times. In their MW satellite sample, the median deviation is smaller than the values we obtain, perhaps reflecting their more detailed modeling of the MW host potentials. 

Our results are also consistent with the conclusions of \citet{2022MNRAS.512..739D}, who studied satellite orbit reconstruction in the ELVIS dark-matter-only simulations. Their orbit model included an NFW host potential and, in some cases, an LMC-like massive satellite, with host and satellite masses evolved using merger-tree information. They found that approximate potentials can produce substantial uncertainties in recovered orbital properties, especially in systems affected by a recent massive accretion event. Our analysis differs in that we focus on LG analog pairs in TNG100 and use static analytic potentials, but we similarly find that massive perturbers and idealized host models can contribute to the scatter in reconstructed histories.

It is also useful to compare this systematic reconstruction error with the uncertainties in recent observational studies of isolated LG dwarfs.~\citet{2024ApJ...971...98B,2025ApJ...993..228B} use present-day 6D phase-space measurements, together with Monte Carlo orbit integrations to determine whether outer LG dwarfs are on their first infall, are backsplash systems, or have previously interacted with the MW, M31, or M33. The orbital uncertainties in these analyses reflect the range of trajectories given observational errors, as well as uncertainties in the adopted LG model, including the masses and present-day phase-space coordinates of the MW, M31, LMC and M33. In contrast, our $\eta$ statistic directly compares reconstructed orbits to the true cosmological trajectories and therefore measures the systematic error associated with the fiducial orbit model itself. The $(\sim 40\%)$ scatter that we find by 6 Gyr is therefore in addition to the observational uncertainties used in these analyses. Future Gaia and HST observations are expected to reduce the observational error so that in the future the errors will be systematics dominated.
 
The main difference between our work and previous single-host studies is the dynamical regime being tested. Outer LG satellites probe the transition region between the virial volumes of the two dominant hosts and the larger LG environment. Their orbits are therefore sensitive not only to the internal structure of one host, but also to the relative motion of the primary-secondary pair and the time-dependent tidal field of the LG analog. Despite this additional complexity, the reconstruction errors are comparable to those found in previous studies. This suggests that simplified two-host orbit models can provide a useful first-order description of outer LG satellite dynamics, while also highlighting the need for caution when interpreting individual orbital histories several Gyr in the past.

\begin{deluxetable*}{lcccccc}
\tablecaption{Massive outer satellites in Local Group analogs. We define a satellite as massive if its total \texttt{SUBFIND} mass exceeds 10 percent of the mass of the less massive primary in the corresponding LG pair. Only outer satellites and LG analogs containing at least one massive outer satellite are listed. Outer satellites are farther than 300~kpc from both hosts at $z=0$. Satellite masses are given in units of $10^{10}\,M_\odot$. The final column gives the total number of outer satellites in each LG analog.
\label{tab:massiveSatellites}}
\tablewidth{0pt}
\tablehead{
\colhead{Catalog} &
\colhead{Primary ID} &
\colhead{Secondary ID} &
\colhead{$N_{\rm outer,massive}$} &
\colhead{Massive Outer Satellite ID(s)} &
\colhead{Satellite Mass(es)} &
\colhead{$N_{\rm outer}$}
}
\startdata
TNG100-1 Hydro & 378745 & 378746 & 1 & 378747 & 41.439 & 55 \\
TNG100-1 Hydro & 387067 & 387068 & 1 & 387069 & 67.739 & 88 \\
TNG100-1 Hydro & 407606 & 407607 & 1 & 407608 & 17.553 & 75 \\
TNG100-1 Hydro & 417122 & 417123 & 1 & 417124 & 14.499 & 61 \\
TNG100-1 Hydro & 434787 & 524216 & 2 & 434788, 434789 & 24.463, 15.101 & 92 \\
TNG100-1 Dark  & 547288 & 547289 & 1 & 547292 & 36.821 & 130 \\
TNG100-1 Dark  & 599434 & 707348 & 2 & 599435, 599436 & 25.571, 16.253 & 125 \\
TNG100-1 Dark  & 615377 & 615378 & 2 & 615379, 615380 & 28.467, 26.612 & 30 \\
TNG100-1 Dark  & 657694 & 739080 & 1 & 657695 & 9.225 & 32 \\
\enddata
\end{deluxetable*}

\begin{table*}[ht]
\centering
\caption{Comparison of the simulations and orbit-modelling assumptions in this work and previous studies.}
\label{tab:model_comparison}
\renewcommand{\arraystretch}{1.25}
\setlength{\tabcolsep}{3pt}
\scriptsize

\begin{tabular}{llll}
\hline
\parbox[t]{0.15\textwidth}{\raggedright\textbf{Property}} &
\parbox[t]{0.25\textwidth}{\raggedright\textbf{This work}} &
\parbox[t]{0.25\textwidth}{\raggedright\textbf{\citet{2024MNRAS.527.8841S}}} &
\parbox[t]{0.25\textwidth}{\raggedright\textbf{\citet{2022MNRAS.512..739D}}} \\
\hline

\parbox[t]{0.15\textwidth}{\raggedright Simulation} &
\parbox[t]{0.25\textwidth}{\raggedright TNG100-1 Hydro and TNG100-1-Dark} &
\parbox[t]{0.25\textwidth}{\raggedright FIRE-2 hydrodynamical zoom-in simulations} &
\parbox[t]{0.25\textwidth}{\raggedright ELVIS dark-matter-only zoom-in simulations} \\

\parbox[t]{0.15\textwidth}{\raggedright Physics included} &
\parbox[t]{0.25\textwidth}{\raggedright Hydrodynamical and dark-matter-only simulations} &
\parbox[t]{0.25\textwidth}{\raggedright Hydrodynamical simulations} &
\parbox[t]{0.25\textwidth}{\raggedright Dark-matter-only simulations} \\

\parbox[t]{0.15\textwidth}{\raggedright Host systems} &
\parbox[t]{0.25\textwidth}{\raggedright Local Group analog pairs with two massive hosts} &
\parbox[t]{0.25\textwidth}{\raggedright Milky Way/M31-mass host analogs} &
\parbox[t]{0.25\textwidth}{\raggedright Milky Way-mass host analogs, including systems with an LMC-like satellite} \\

\parbox[t]{0.15\textwidth}{\raggedright Mass resolution} &
\parbox[t]{0.25\textwidth}{\raggedright \(m_{\rm DM}\sim 7.5\times10^6\,M_\odot\) in TNG100-1; \(m_{\rm DM}\sim 8.9\times10^6\,M_\odot\) in TNG100-1-Dark; \(m_{\rm baryon}\sim 1.4\times10^6\,M_\odot\)} &
\parbox[t]{0.25\textwidth}{\raggedright \(m_{\rm DM}\sim 3.5\times10^4\,M_\odot\); \(m_{\rm star}\sim 5\times10^3\,M_\odot\)} &
\parbox[t]{0.25\textwidth}{\raggedright \(m_{\rm DM}\sim 1.9\times10^5\,M_\odot\)} \\

\parbox[t]{0.15\textwidth}{\raggedright Satellite sample} &
\parbox[t]{0.25\textwidth}{\raggedright Subhaloes with \(M_{\mathrm{total}}>1\times10^8\,M_\odot\) and \(r > 300\,\mathrm{kpc}\) at \(z=0\) from both hosts} &
\parbox[t]{0.25\textwidth}{\raggedright Stellar-mass selected satellites with \(M_\star>3\times10^4\,M_\odot\)} &
\parbox[t]{0.25\textwidth}{\raggedright Subhaloes with \(M_{\rm peak}>10^9\,M_\odot\)} \\

\parbox[t]{0.15\textwidth}{\raggedright Orbit model} &
\parbox[t]{0.25\textwidth}{\raggedright Static analytic potentials with NFW halos; Hydro hosts also include an approximate disk component. Massive satellites are included as Plummer spheres in orbit integrations} &
\parbox[t]{0.25\textwidth}{\raggedright Static axisymmetric potentials fitted to the \(z=0\) host mass profiles, with generalized NFW halos and double-exponential disks} &
\parbox[t]{0.25\textwidth}{\raggedright Spherical NFW host potential; systems with an LMC-like satellite also include a spherical Hernquist satellite potential} \\

\parbox[t]{0.15\textwidth}{\raggedright Time dependence} &
\parbox[t]{0.25\textwidth}{\raggedright Host masses fixed during backward integration} &
\parbox[t]{0.25\textwidth}{\raggedright Host masses fixed during backward integration} &
\parbox[t]{0.25\textwidth}{\raggedright Host and massive-satellite masses evolved using merger-tree information} \\

\parbox[t]{0.15\textwidth}{\raggedright Dynamical friction} &
\parbox[t]{0.25\textwidth}{\raggedright Included} &
\parbox[t]{0.25\textwidth}{\raggedright Not included} &
\parbox[t]{0.25\textwidth}{\raggedright Included} \\

\hline
\end{tabular}
\end{table*}

\section{Conclusion}
\label{sec:summary}

We have used the \textsc{IllustrisTNG} simulations to test how accurately backward orbit integrations in static potential models recover the histories of satellites in Local Group analog systems. Our sample consists of MW--M31-like pairs selected from TNG100-1 Hydro and TNG100-1-Dark using cuts on host mass, separation, relative velocity, mass ratio, and isolation. Around each analog, we identify satellites with total bound \texttt{SUBFIND} mass \(M_{\rm total}>10^8\,M_\odot\), and focus in particular on outer satellites that lie beyond \(300\,\mathrm{kpc}\) from both hosts at \(z=0\). For each satellite, we compare the true trajectory extracted from the simulation merger trees to a reconstructed trajectory obtained by integrating the present-day phase-space coordinates backward for \(6\,\mathrm{Gyr}\) using \texttt{gala}.

Our main conclusions are as follows. First, simplified orbit integrations recover the broad orbital evolution of outer LG satellites, but the reconstruction error increases with lookback time. The reconstructed and simulated trajectories agree best at recent times, while both the median fractional distance error and the \(1\sigma\) scatter grow toward \(t_{\rm lb}\sim 6\,\mathrm{Gyr}\). This trend is consistent with the expectation that backward integrations become increasingly sensitive to neglected time dependence, halo structure, and environmental perturbations.

Second, the two dominant hosts provide the leading contribution to the dynamics of the outer-satellite population. Even when the primary and secondary are modeled as point masses, the reconstructed orbits broadly track the simulated trajectories. However, the point-potential model produces larger median offsets and larger scatter than the fiducial orbit model, demonstrating that the internal mass distribution of the hosts is important for improving the accuracy of individual satellite orbits.

Third, the inclusion of an approximate baryonic disk in the Hydro host potentials does not significantly change the reconstruction accuracy for the outer satellites considered here. The Hydro integrations with NFW halos only and those with NFW halos plus disks produce similar median errors and scatter. This suggests that, for this outer-satellite sample, the dominant uncertainties are not driven by the disk component alone, but instead by the simplified treatment of the full LG environment.

Fourth, massive satellites can affect individual systems but do not appear to dominate the reconstruction error across the full sample. In systems containing companions with masses exceeding ten percent of the less massive host, we compared three-body integrations to integrations that include the additional massive bodies. The resulting orbits are generally similar, indicating that massive satellites are a secondary source of scatter for the population as a whole, though they may be more important for particular analogs, depending on the orbital configurations of the massive satellites.

Finally, the residual mismatch between reconstructed and simulated orbits is likely produced by a combination of physical effects: host mass growth, halo asphericity, close passages with the hosts. We do not find evidence for a single dominant source of error that explains all outliers. Instead, the accuracy of a reconstructed orbit depends on the detailed dynamical history of each host environment.

Overall, our results support the use of backward orbit integrations in multi-host orbit models as a practical tool for interpreting the present-day phase-space coordinates of outer LG galaxies. Such models are most reliable for recovering broad statistical trends and recent orbital evolution. However, individual orbital histories become increasingly uncertain at earlier lookback times, especially beyond \(\sim 4\)--\(5\,\mathrm{Gyr}\), and classifications that depend on precise pericenter times or distances should be interpreted with this systematic uncertainty in mind. Future work that incorporates time-dependent host growth, non-spherical potentials, and constrained Local Group environments will be needed to improve orbit reconstruction for individual outer Local Group galaxies.

\begin{acknowledgements}
LES is supported by the U.S. DOE Grant DE-SC0010813. EP acknowledges support from HST GO-17501. Support for GO-17501 was
provided by NASA through a grant from the Space Telescope Science Institute, which is operated by the Association of Universities for Research in Astronomy, Inc., under NASA
contract NAS 5-26555. 
\end{acknowledgements}
\begin{software}
{Astropy} \citep{2013A&A...558A..33A,2018AJ....156..123A,2022ApJ...935..167A},
{Gala} \citep{2017JOSS....2..388P},
{NumPy} \citep{2020Natur.585..357H},
{SciPy} \citep{2020SciPy-NMeth},
{Matplotlib} \citep{2007CSE.....9...90H}, and
{pandas} \citep{2010scpy.soft.....M}.
\end{software}

%% Bibliography
\bibliographystyle{aasjournalv7}
\bibliography{references}

\end{document}